# *"It answers questions that I didn't know I had"*: Ph.D. Students' Evaluation of an Information-Sharing Knowledge Graph

Stanislava Gardasevic, University of Hawaii at Manoa
Manika Lamba, University of Illinois at Urbana-Champaign

**Purpose:** Interdisciplinary Ph.D. programs can be challenging as the vital information needed by students may not be readily available; it is scattered across the university's websites, while tacit knowledge can be obtained only by interacting with people. Hence, there is a need to develop a knowledge management model to create, query, and maintain a knowledge repository for interdisciplinary students. We propose a knowledge graph containing information on critical categories (faculty, classes, dissertations, etc.) and their relationships, extracted from multiple sources, essential for interdisciplinary Ph.D. students. This study evaluates the usability of a participatory-designed knowledge graph intended to facilitate information exchange and decision-making.

**Methodology:** We used data from multiple sources (such as university websites, faculty profiles, publication and dissertation metadata, and crowdsourced data) to generate a knowledge graph in the Neo4J Bloom platform. We recruited 15 interdisciplinary Ph.D. students using convenience sampling from the University of Hawaiʻi at Mānoa at various Ph.D. stages to design and populate the knowledge graph. Next, we conducted a mixed methods study to perform its usability evaluation. Firstly, we engaged the students in a participatory design workshop to identify relevant graph queries. Secondly, we conducted semi-structured interviews to determine the usability of the knowledge graph and rate the queries. Each interview was coded with structural and thematic codes and was further analyzed using sentiment analysis in R programming language.

**Findings:** The usability findings demonstrate that interaction with this knowledge graph benefits Ph.D. students by notably reducing uncertainty and academic stress, particularly among newcomers. Knowledge graph supported them in decision-making, especially when choosing collaborators (e.g. supervisor or dissertation committee members) in an interdisciplinary setting. Key helpful features are related to *exploring student-faculty*

*networks, milestones tracking, rapid access to aggregated data,* and *insights into crowdsourced fellow students' activities*. However, they showed concerns about crowdsourced data privacy and accessibility. Although participants expressed the need for more qualitative data in the graph, they noted it helped identify people to talk to about the topics of their interest.

**Originality:** The knowledge graph provides a solution to meet the personalized needs of doctoral researchers and has the potential to improve the information discovery and decision-making process substantially. It also includes the tacit knowledge exchange support missing from most current approaches, which is critical for this population and establishing interdisciplinary collaborations. This approach can be applied to other interdisciplinary programs and domains globally.

**Introduction**

Obtaining a Ph.D. can be perceived as a self-discovery journey, promising the potential of a rewarding academic career and freedom (Wood et al., 2016). But for many, graduate studies can be very stressful and strenuous (Bair and Haworth, 2004; Grady et al., 2014). A significant percentage of Ph.D. students have reported experiencing anxiety and depression (Levecque et al., 2017; Woolston, 2019), loneliness, or imposter syndrome (Chakraverty, 2020). Also, the degree may take a long time to obtain. For example, in the US, the median range to complete Ph.D. is 6-12 years, depending on the discipline (National Center for Science and Engineering Statistics, 2021), while the dropout rate is estimated at a disturbing 50% (Cassuto, 2013). Success factors are often related to navigating the dissertation process (Young et al., 2019) and the so-called "departmental culture", including student/faculty relationship, student satisfaction with the program, student-to-student interactions (Bair and Haworth, 2004) including peer mentorship (Preston et al., 2014). The problem with navigating "departmental culture" exacerbates in interdisciplinary PhD programs (such as those that involve faculty from multiple departments, as in the case of iSchools), since students have to "span boundaries between areas, departments, and knowledge bases" (Gardner, 2011).

This paper details the development and evaluation of a knowledge graph that tackles those issues by answering the information needs of the Ph.D. student population, allowing them to access and share crucial, often-tacit knowledge and identify potential community connections.

Ph.D. students need support to make well-informed decisions when selecting a supervisor and collaborators, and to navigate the program and dissertation process. Still, not all necessary information can be found online. The most pertinent knowledge exchange happens through in-person conversations, serendipitously (Twidale et al., 1997), and is often referred to as "social" information sharing (Talja, 2002). Previous research indicates that women, first-generation college students, and underrepresented minority students may have a stronger need to obtain knowledge about "the unwritten, unofficial, and often unintended lessons, values, and perspectives that make up the "hidden curriculum" in graduate education" (Wood et al., 2016), but have less access to social networks where this knowledge circulates.

This study engaged participatory approaches to involve students in an interdisciplinary Ph.D. program in the design, development, and evaluation of a *knowledge graph* created to address the above-mentioned issues. Knowledge graph is the core of a potential information system intended to democratize access to insider (tacit) knowledge needed for Ph.D. students to complete their degrees successfully, especially for those who don't have equal access to the networks where the knowledge exchange occurs. A knowledge graph model and dataset (previously published) were developed to help students with decision-making on each of the important steps in the program.

Our research followed the recommendations of the ISO (2019) standard for human-centered design principles and activities for enhancing computer-based interactive systems. The previous stages encompassed the requirements gathering study and the design and creation of the knowledge graph. This study presents the evaluation segment, where Ph.D. students reflect on the usefulness of the dataset by utilizing it to answer both specific queries and engage in exploratory searches.

*Case Background - The Interdisciplinary Ph.D. Program*

This research is situated in the Interdisciplinary Ph.D. Program in Communication and Information Sciences (CIS) at the University of Hawaiʻi at Mānoa (UHM). The program has 20-30 students at any given point in time (with over 100 alumni) and includes faculty members from four units across three colleges: *Social Sciences*, *Natural Sciences*, and the *Shidler College of Business*. Even though there are over 40 faculty members appointed to the CIS graduate faculty, only the "core" faculty are interested in interdisciplinary topics such as social or health informatics, information policy, human-computer interaction, etc. The degree requirements for this program include:

i) attending weekly interdisciplinary seminars;

    [first stage]

ii) taking three core courses;

iii) passing three comprehensive exams;

iv) publishing a paper under the supervision of one of the CIS-affiliated faculty;

    [pre-proposal stage]

v) selecting a dissertation research supervisor and four committee members;

vi) defending a dissertation proposal;

    [all-but-dissertation or "ABD" stage]

vii) defending the dissertation.

CIS Ph.D. students have the freedom to choose among a myriad of research avenues which creates a challenge for each student to carve their own path through four different "departmental cultures" (Bair & Haworth, 2004), as well as program requirements taken as the blueprint for the graph design. The official program website suggests the five-year timeline for its completion.

*CIS Knowledge Graph and Related Work*

In our previous work, we engaged community members via a series of interviews and workshops to develop the knowledge graph model representing essential categories, relationships, and data relevant to an interdisciplinary Ph.D. student population (Figure 1). This knowledge graph is a multilayered/multiplex network, representing *People* through 12 layers or dimensions.

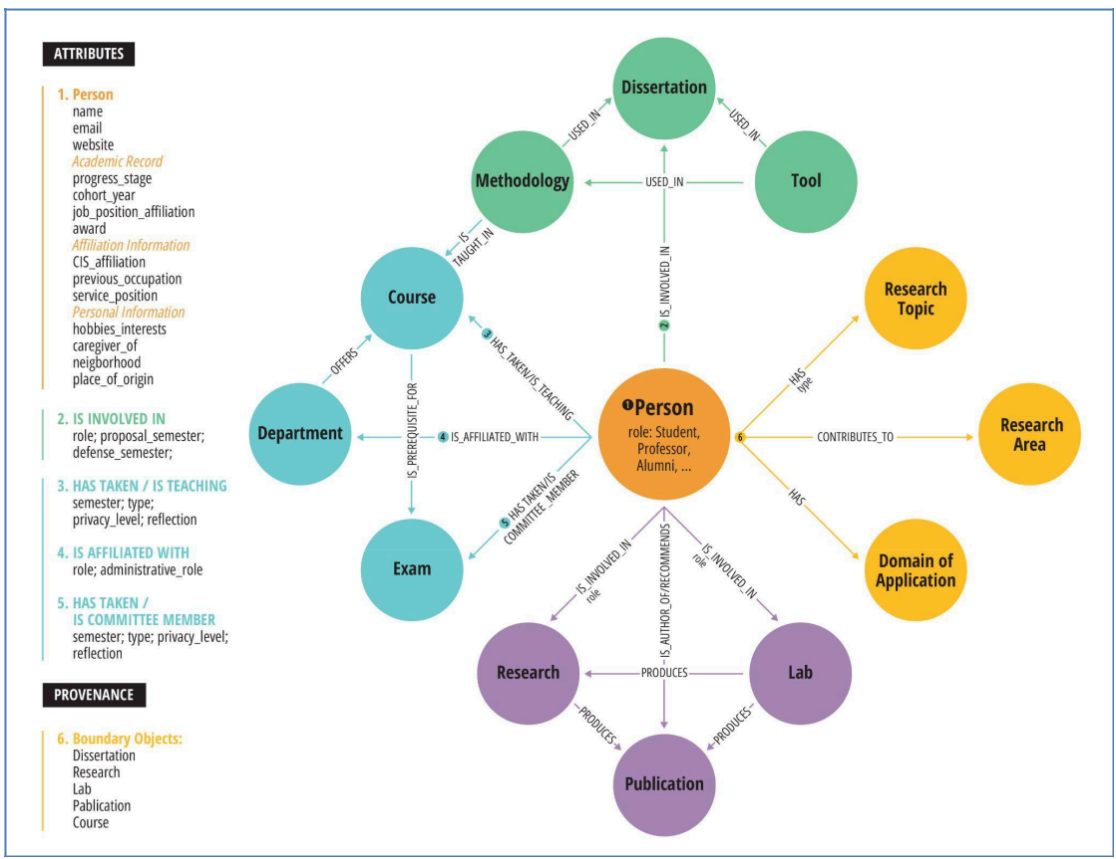

**Figure 1**: CIS Knowledge Graph Model (Gardasevic and Gazan, 2023)

Most recommender systems research focuses on scholarly networks that cover relatively few dimensions, mainly based on publication metadata (e.g., Ortiz Vivar et al., 2022). Knowledge graphs in educational settings have been similarly limited in their dimensions of interest. For example, research by Rahdari et al. (2021) proposes a recommender system matching prospective graduate students with faculty who have expertise in an area and are affiliated with a particular school/department. Research by Jordão et al. (2014) focuses on visualizing patterns elicited from course-taking activity. The KCUBE project developed a knowledge graph to support advising students in computer science regarding curriculum and career planning (Li et al., 2022). However, to understand the full complexity of social structures, researchers suggest that three dimensions are considered minimum (Dickison et al., 2016). Considering scholars' many hats (e.g., teaching, mentoring, researching, publishing, administrating science, etc.) (Börner, 2010), the graph produced for this study attempts to comprehensively represent these activities through the prism of Ph.D. students' information needs.

Due to the current state of the art in the area of automatic tagging (Banerjee et al., 2022), much of the data in the CIS graph was manually curated and captured with the Neo4J graph database. The refining attributes in this graph are intended to support not only information discovery and decision-making but also social connection-building and information sharing in the community, as knowledge seekers are hesitant to ask for information unless they feel they have a close enough relationship with those who could share it (Keppler and Leonardi, 2023).

To support this feature, the system depends on crowdsourced information combined with the information available across the web and metadata on papers/dissertations produced by actors in the local domain. The sources and provenance of the data aggregated in this graph are shown in Figure 2.

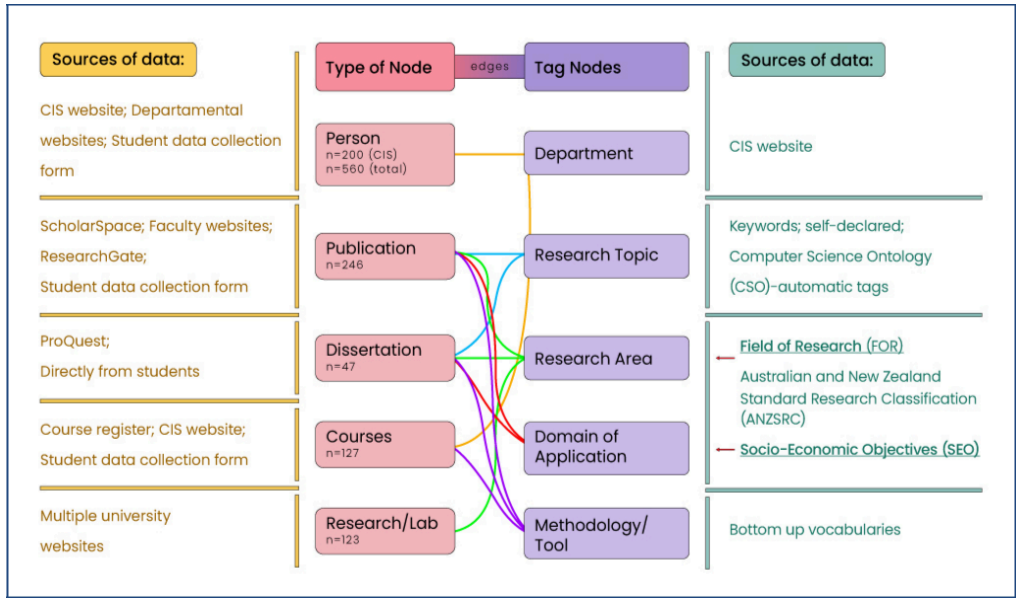

**Figure 2**: Sources of data for CIS Knowledge Graph

**Relevant literature**

*Knowledge Graphs as Medium for Information Discovery*

The research on the dynamics of multidimensional social and knowledge networks is a promising avenue, especially for facilitating transdisciplinary collaborations (Contractor, 2009). Knowledge graphs are often used for knowledge organization and developing systems to support scientific information discovery and expert recommendation (e.g., Callahan et al., 2023; Osborne et al., 2013).

Another affordance of the graph representation of local social connections is the possibility to identify the people in one's domain who can offer direct advice - especially other students who are invaluable sources of information, yet traditionally excluded from expert recommender systems (Wood et al., 2016). Preston et al. (2014) outlined that "peer mentorship is an under-utilized resource with great capacity to foster human and social capital within and between cohorts of graduate students."

Graph-based knowledge organization systems have tended to be monodisciplinary (Osborne et al., 2013), aiming to represent scholars through their universal set of activities (Corson-Rikert et al., 2012), which can be applied to local settings. In this study, we were inspired by the Hawaiian epistemological approach to knowledge (Meyer, 2003), where "*knowing*" can not be separated from the physical place where it emerged and depends on one's relationships and lineage, providing a bottom-up approach to knowledge organization system creation.

*Knowledge Graph to Support Interdisciplinary Science and Education*
Interdisciplinary science is becoming increasingly common (Porter & Rafols, 2009), more successful at making breakthroughs, and generating more relevant outcomes than monodisciplinary research (Fortunato et al., 2018). Even though the CIS knowledge graph model was developed to satisfy the specific needs of the Ph.D. students in a particular interdisciplinary program, the categories and relationships, as declared in the knowledge graph schema, can be applied to other academic domains and settings.  For example, it can be applied to facilitate interdisciplinary or multidisciplinary research in fields such as information science/iSchools (Wiggins and Sawyer, 2012), life sciences (Franc-Dąbrowska et al., 2020), or climate change research (Bruine de Bruin and Morgan, 2019).

***Knowledge graph Evaluation Related Work***
Knowledge graph evaluation methods, most commonly utilizing a probabilistic approach, deal with knowledge graph completion and error detection, and the research is mainly focused on DBpedia, which is the most prominent knowledge graph (Paulheim, 2016). Previous research by Sarrafzadeh et al. (2014) employed mixed methods to examine general users' exploratory search behavior when interacting with knowledge graphs and pertinent corresponding documents based on Wikipedia articles; the authors consequently developed so-called hierarchical knowledge graphs (Sarrafzadeh et al. 2017; 2020) that were evaluated via experimental approach. Our contribution to this body of research is methodological- i.e., engaging the intended end users in a qualitative evaluation study of a knowledge graph that was created based on their information needs. Even though qualitative approaches are applied to gather requirements and evaluate visual analytical tools to support decision-making in other areas (Dimara et al., 2022; Hong et al., 2020), there is a lack of literature that uses such an approach to investigate the end users' experiences with knowledge graph tools, which is critical when it comes to their adoption (Li et al., 2023).

**Research Questions**

The research questions guiding this study were formulated to evaluate the usefulness and potential impact of the knowledge graph created based on extensive requirements gathering study and inputs of end users.

**RQ1-** What are the reactions and impressions of CIS Ph.D. students to the knowledge graph?

**RQ2-** Which knowledge graph features are perceived as helpful? How do students talk about their next steps in the Ph.D. program upon interacting with knowledge graph?

**RQ3-** What information was missing from the knowledge graph?

**Methodology**

The entire methodological approach as conducted in the overall study is shown in Figure 3. The study was divided into three stages, pertinent to human-centered design approach, and different data collection methods were applied in each stage. In the (i) requirement gathering stage, we conducted interviews and the program website usability study with PhD students and  (omitted); in the (ii) graph design and data ingestion stage, we conducted three workshops, scraped the web, indexed publications and dissertations, and collected and ingested the crowdsourced data into Neo4J graph database (omitted). The methods conducted in this study's final iii) evaluation stage are described in detail in the following text, and they encompass the participatory design workshop, a survey, and semi-structured interviews with CIS PhD students.

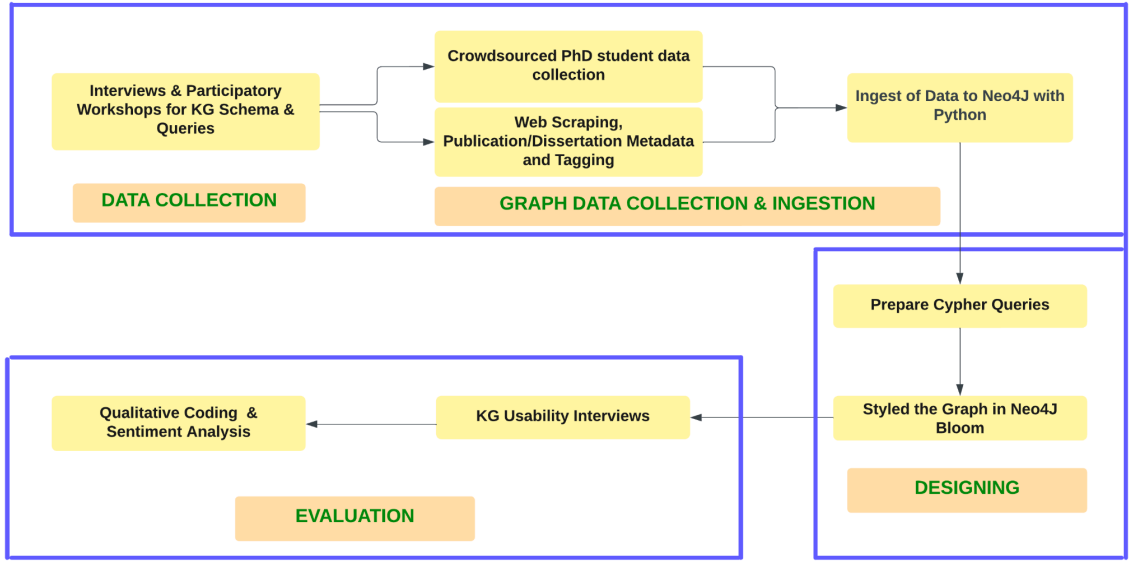

**Figure 3**: Summary of the Methodological Approach

*Data collection*

Our data collection approach in this phase was three-fold: (i) a participatory design workshop ; ii) follow-up survey; and (iii) in-depth interviews with CIS students following the knowledge graph evaluation protocol. These approaches were chosen to gather and address both group and individual insights about the proposed technology. The follow-up survey allowed the participants to voice their opinions asynchronously and anonymously.

The interdisciplinary CIS PhD students were invited to participate in this study via CIS student mailing list and were asked to sign the consent form. All of the data was collected by the first author.

*Participatory design workshop*

In June 2022, a participatory design workshop convened, drawing the active engagement of **15 Ph.D. students**, split between in-person (n=8) and online (n=7) participation. The workshop was concurrently scheduled with the group's weekly interdisciplinary seminar, providing a convenient opportunity for students to participate without making it mandatory. The participants were not offered any monetary or other incentives.

The workshop agenda included a preview of the knowledge graph model, specific queries, and a community discussion on potential system functionalities. Furthermore, students were actively involved in two distinct voting exercises: one focusing on refining the multiplex graph visualization design, and the other prioritizing useful queries for this population. The students were invited to suggest other specific questions they found significant.

Follow-up survey

The workshop was followed up by the post-workshop survey (in Google Forms) that was distributed to the CIS student mailing list, and contained several open and closed-ended questions evaluating the usefulness of the graph demonstrated. Some survey results **(n=8)** showed that participants perceived the tool as useful. To investigate further which features students found useful, we developed a graph evaluation protocol that addresses the research questions.

*Semi-Structured Interviews and Knowledge Graph Evaluation Protocol*

In spring 2023, the authors interviewed **eight** CIS PhD students (38% of the population) using the *graph evaluation protoco*l [1]. We applied convenience sampling and the first author directly emailed potential participants to invite them to participate in the study. Each interview was 82 minutes long on average, and participants were offered a $50 gift card [3] for their partaking. We used a QuickTime player to screen and voice record each interview and the interactions with the graph.

Even though the graph was primarily designed to support new students and their information needs, we invited students from different stages in the interdisciplinary CIS PhD program to evaluate it. Among study participants four were from the latest cohort (or "new" students); two were in the second year of the program (or "mid-stage" students), and two were in their finishing stage ("all-but-dissertation" or ABD students). Of this population, six were domestic/US students, and two were international. We stopped interviewing new participants as we noticed a thematic saturation (Saunders et al., 2018) that emerged from the data collected from participants in each of the stages.

We used the Neo4J Bloom graph exploration tool, its styling capabilities, and natural language querying options to showcase and evaluate the CIS knowledge graph . Led by the results from the participatory design workshop, we created a list of 20 queries voted as useful at least three times (Appendix 1) that were the crux of the *graph evaluation protocol*. The *protocol* encompassed (i) identifying categories of students' interest in the controlled vocabulary entries [2] used to tag *Publications* and *Dissertations*; (ii) asking for current information needs and exploring the graph to seek answers; (iii) rating each of the 20 queries on a five-point Likert scale, with occasional prompts to reflect on their ratings; ending with (iv) five summative questions.

Three interviews were conducted in person, during which students engaged in hands-on exploration of the graph in two instances. Throughout all the interviews, the researcher played the intermediary role in guiding the participants through the graph exploration exercises. This was perceived appropriate considering the complexity of the interface tool (Neo4J Bloom) and the fact that the researcher provided an "*informal advising*" session that accompanied the evaluation protocol, sharing her views, advice, and experiences related to the program, as she was demonstrating the graph queries and facilitating the exploration.

Figure 4 shows an example of the interface, demonstrating the exploration of a particular faculty member and the *research areas* he publishes in (yellow nodes). Such queries were often expanded to show other listed categories based on participants' interests.

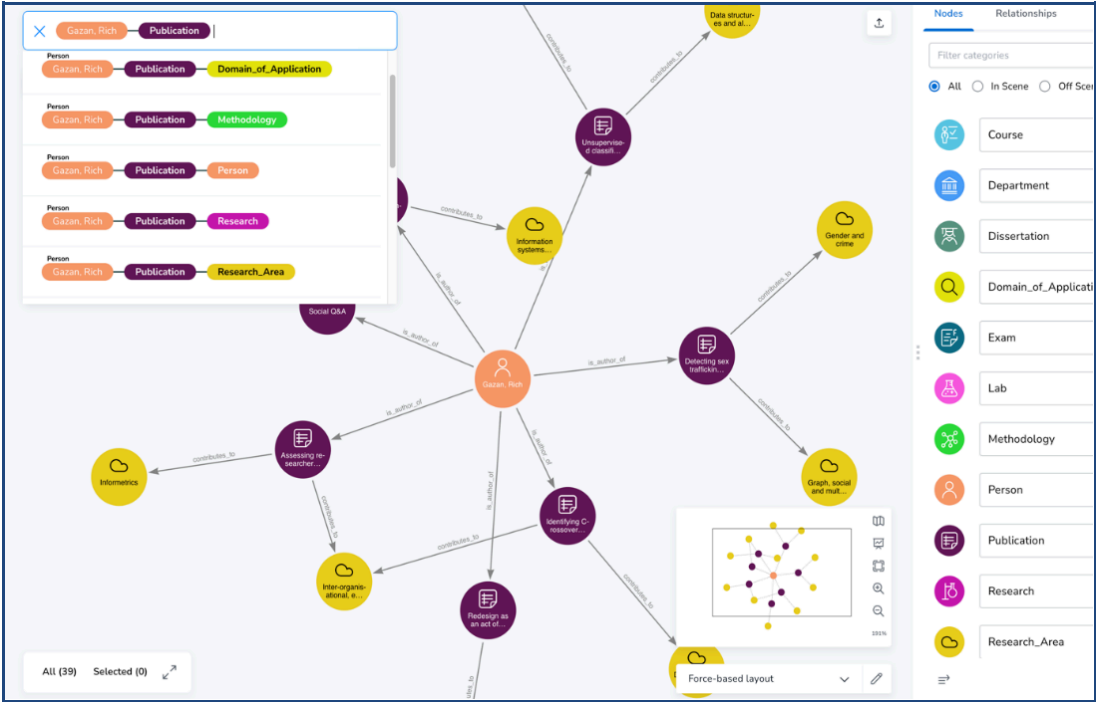

**Figure 4:** Example of the Knowledge Graph interface in Neo4J Bloom

We were not evaluating Neo4J Bloom's usability, but rather the information that can be obtained from the knowledge graph.

*Data Analysis*

The interviews were transcribed using the Otter.ai [4] transcription tool, after which both authors revised the automatically generated text for accuracy. The transcriptions were anonymized, and analyzed in Atlas.ti cloud software. We coded each interview (with structural and thematic codes) and grouped them into overall themes reported in the following section. Structural coding assigns concise phrases to data segments relevant to specific questions of inquiry (Saldaña, 2016), and in our study, these codes correspond to graph categories. Thematic coding involves identifying, analyzing, and reporting patterns or themes within collected data to capture important aspects in relation to the research questions (Braun & Clarke, 2006).

Furthermore, Appendix 2 demonstrates how we conducted sentiment analysis using the *syzuhet* library (Jockers, 2015) in R. This analysis served two purposes: (i) assessing how the narrative was structured to convey sentiments across participants' responses, and (ii) identifying emotions (*disgust, anger, anticipation, surprise, trust, fear, sadness, and joy*) and sentiments (*positive and negative*) within the participants' responses.

*Positionality and Limitations*

The first author is an insider to the program, a Ph.D. student herself, and a study participants' peer. Aside from the obvious advantages of this approach that encompass her familiarity with jargon, social rules, and norms, the drawbacks of the approach have to do with the "courtesy bias" on the participants' side that might have resulted in more favorable ratings, than it would have been if the researcher was an outsider.

The second author is an outsider to the program, yet a recent Ph.D. graduate who went through the same process, just in a different country and culture; her insights serve as an objective check and set of eyes on the data and results.

**Findings**

This section gives an overview of qualitative coding and sentiment analysis results. The codes (reported as **bolded text**) were organized in themes pertinent to the research questions and reported as subsections and sections, respectively. We report the number of code occurrences over all interviews as "*c*."

*Reactions and Impressions of CIS Ph.D. Students to the Knowledge Graph*

Seven out of eight participants found the graph exploration results **interesting** ($c=35$), and four were **surprised** ($c=18$). On the other hand, five participants got results that showed **known information** ($c=10$), while four got to **confirm their beliefs** ($c=9$).

The discussions accompanying the graph exploration tasks called for reflective commentary on the participants' side. To demonstrate the overall flow of conversation and the participants' sentiments during the interview, we refer to Figure 5, for which we used a binning average method (Jockers, 2015) to track changes in polarity in participants' responses. Figure 5 shows that the new students have ended the interview and graph exploration exercise (conditionally taken as an intervention) with a much more optimistic outlook when compared to ABD students.

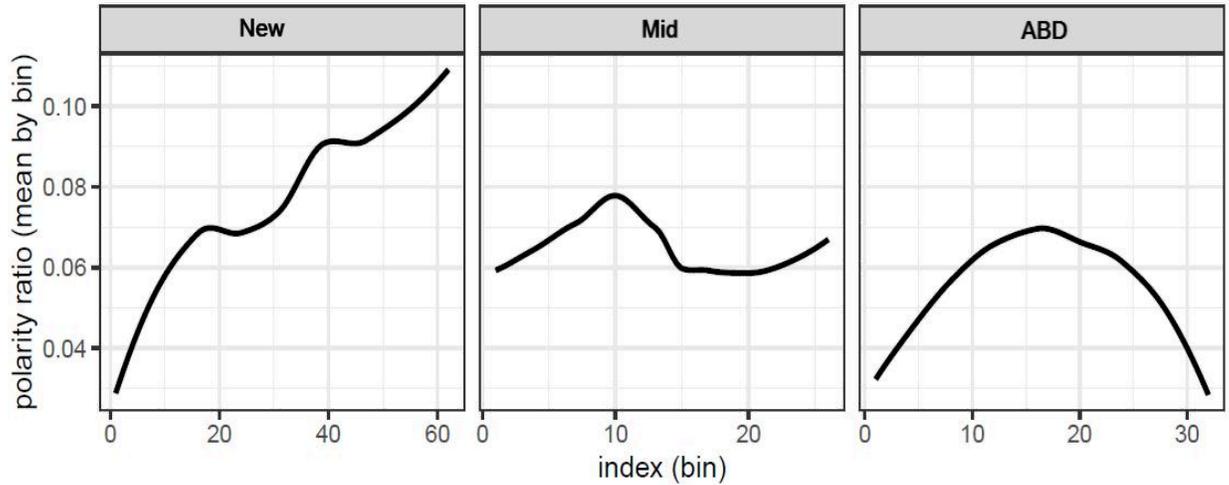

**Figure 5**: Sentiments polarity over the participants' responses

Figure 6 (a-b) illustrates the distribution of emotions and sentiments expressed by the participants during the interview, providing valuable insights into their overall experience of witnessing the knowledge graph in a live demo. The findings reveal a highly positive reception among the participants (Figure 6a). The analysis reveals that *"trust"* emerged as the most prevalent emotion among the participants when the knowledge graph was used for information retrieval. This suggests that the participants had confidence in the knowledge graph's capabilities and reliability to provide accurate and relevant information. Such trust is crucial for successfully adopting and accepting knowledge graph tools (Li et al., 2023). Following *trust*, *"anticipation"* and *"joy"* were among the prominent emotions expressed by the participants (Figure 6b). This finding indicates that the participants were excited about the possibilities offered by the knowledge graph and felt a sense of pleasure or happiness while interacting with it. Further, *"anticipation"* often suggests an eagerness to explore and discover new information through the knowledge graph.

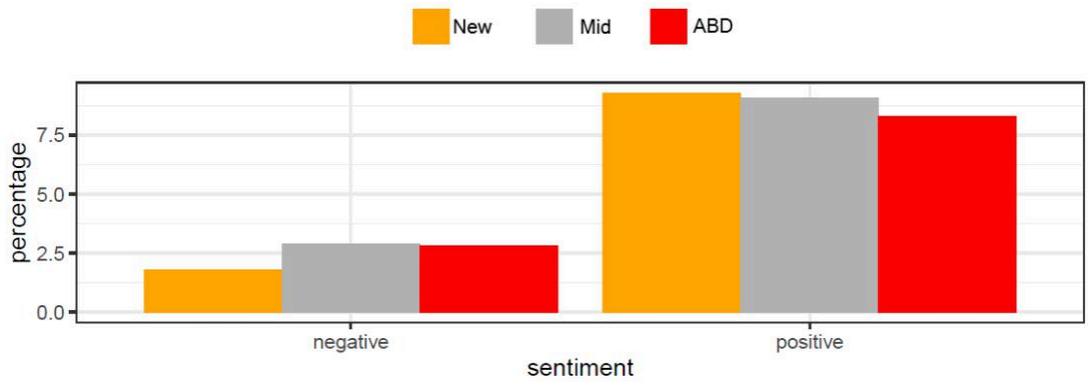

*a) Sentiments*

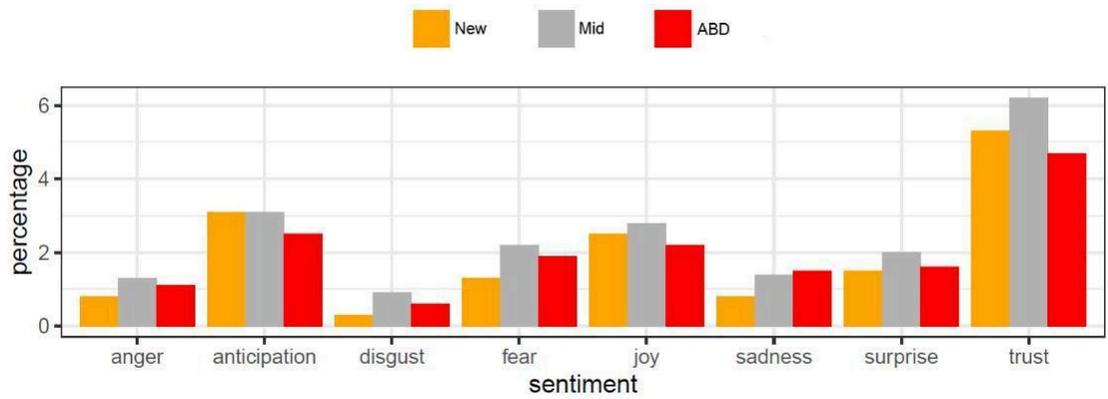

*b) Emotions*

**Figure 6:** Percentage of Emotions and Sentiments in the Participants' Responses

*Different Approaches to the Knowledge Graph Based on the Program Stage*
Students in different stages of the program had different experiences with the graph. That was evident both from the sentiment analysis results, structural coding, and the results from the query ratings (Appendix 1). For example, when participants were asked to start a search on a category of their interest, the new students wanted to learn more about *courses/exams* and *faculty*. In contrast, the second-year students started searches to identify pathways of their *peers* (e.g., where and when they published papers), *faculty* engagements in *dissertations*, or research-related questions (e.g., exploring *methods* of their interest). The *faculty research and publishing* information was considered more valuable to the students closer to finishing the degree and conducting research themselves.

*Ph.D. Life Hardships and Peer Support via Knowledge Graph.* The two most prominent difficulties pertinent to the CIS program are its interdisciplinary nature and the lack of support. Following is one reflection on the program organization, where a new student lacked guidance when they first started the program:

> *It was a bit disorganized, as a new student was not sure what I was doing. So much new information.*

Even though the guidelines clearly indicate the expected milestones for Ph.D. students, many newcomers experience uncertainty, even "imposter syndrome," in the first year of their program. They benefit from hearing experiences and tips from those who have already gone through them; still, such exchange doesn't happen often in this program. One student described their Ph.D. journey as being a "super lonely path,' especially in an interdisciplinary program such as CIS, where there are no labs or student collaborations:

> *In the CIS, you are doing your own thing. Sometimes people would have no idea what you're doing except for you and your supervisor.*

Thus, students valued the access to information crowdsourced from their peers, which served as motivation for some or as a comforting factor for others- a reinforcement that they are on the right path.

*Helpful Features of CIS Knowledge Graph*

Throughout the interviews, the words **'helpful'** and **'useful'** were explicitly used over 50 times to refer to different aspects of the knowledge graph. The **whole graph exploration experience**, including the visualization of connections among people and concepts in the graph, was perceived as the most helpful, compared to any individual feature. A 'mid-stage' student commented on that with:

*A lot of people who are… part of these types of networks probably aren't even aware of who's in this whole network. So, for you to have kind of a bird's eye view of who's all involved, helps you to be able to take the right steps … towards that problem-solving.*

The following text outlines specific themes participants found helpful when interacting with the knowledge graph.

*Previous students' activities.* Information that was found very useful by most Ph.D. students was the data on the activities of their peers and predecessors. Newer students who needed to plan where to publish papers and consider the faculty members for collaboration found the queries around *student publications* and their *co-authorships with faculty* informative, considering there is no easy way to access this data currently. Figure 7 shows the screenshot of a part of the co-authorship query in which a student could recognize which faculty members publish with their students (orange and yellow nodes) vs. those who work with other faculty in the same department (same color nodes) or have cross-departmental collaborations.

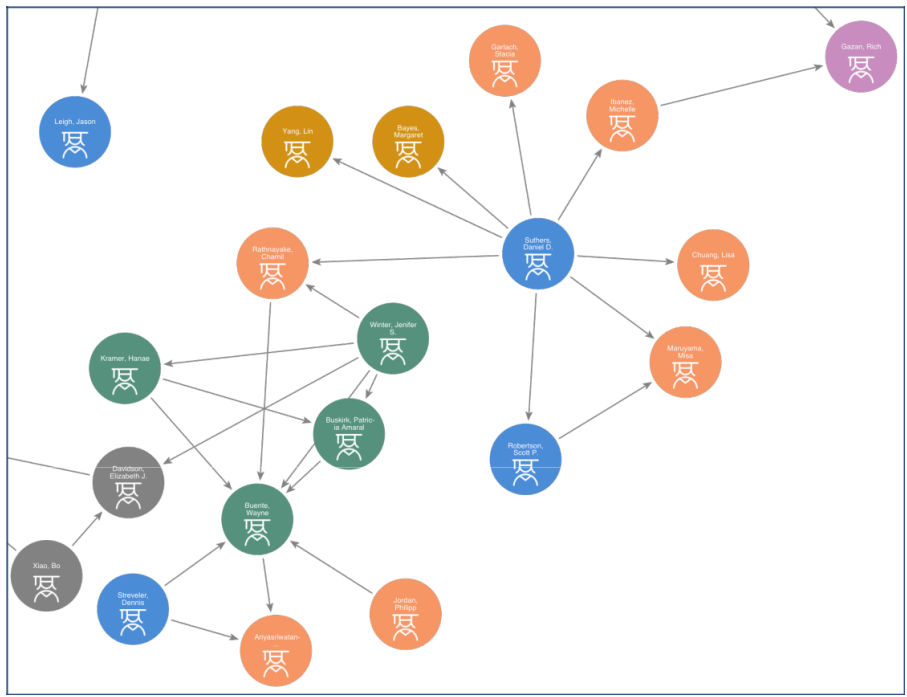

**Figure 7:** A part of the co-authorship query that depicts only the CIS actors

Similarly, participants appreciated the ability to discover and explore **previous students' dissertations**, their committees, and other aspects, such as *methods* and *tools* used to perform research. For example, Figure 8 shows the screenshot of a query where a participant looked at dissertations in the *area* of "Collaborative and Social Media" and the research methods used in them. This query was expanded to investigate the dissertation of interest further by showing its *research topic* and *domain of application*.

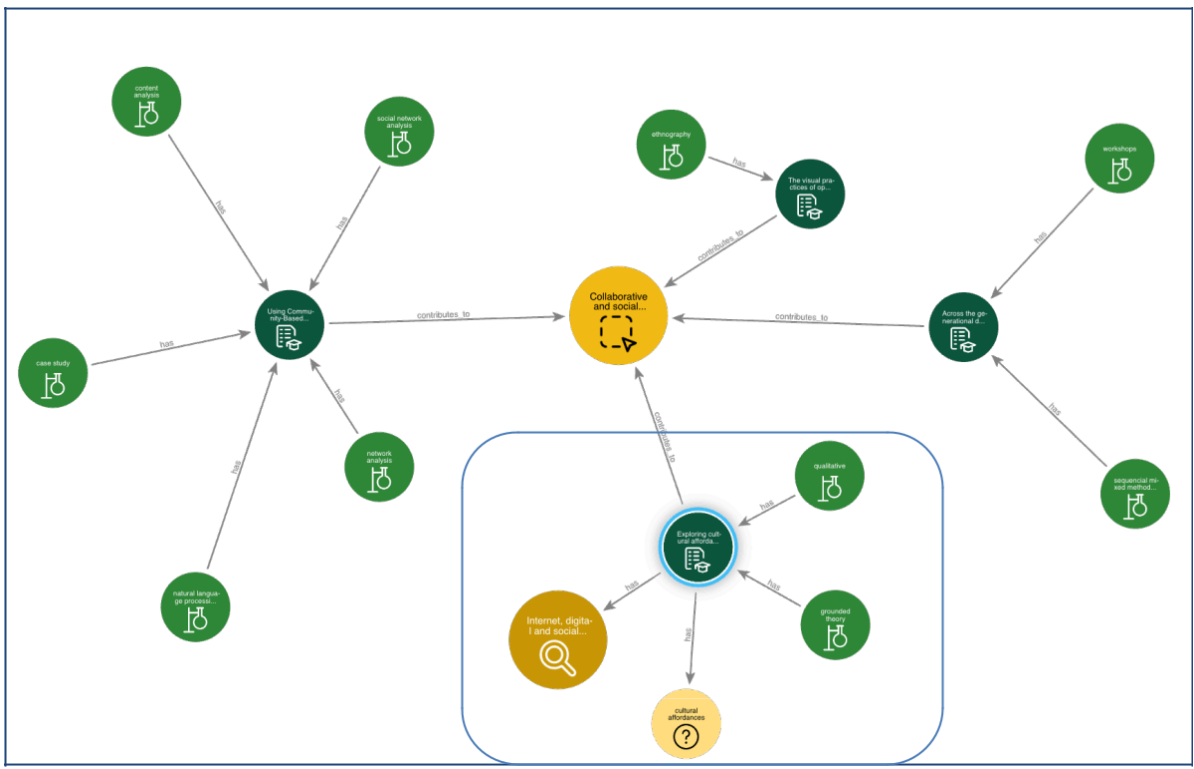

**Figure 8**: Example of a query exploring previous students' dissertations

*Student Reflections on Courses/Exams.* Seven out of eight participants found the **students' reflections on classes** useful. A new student commented on this feature:

> *When you go through the class availability, and all of the hundreds of frickin' departments;... it's really overwhelming to figure out what type of courses are useful... I think something like this would show me- other people in our department took these types of courses- here are some that might be helpful.*

Similarly, **students' reflections on "directed readings"** with faculty members gave new students insights into individual professors' supervision styles. This information was found particularly useful, even critical, next to other aspects that can support students' **decision-making** when it comes to picking the supervisor or committee members:

> *The other piece that was helpful was just reading feedback from other peers on the selection of their chairs or working with the chairs. It's one lens, one perspective, but it's a data point for me to consider. So, that brings me just kind of awareness before I make the decision.*

*Faculty Exploration.* Exploration of **people** was one of the features that students found very useful, especially when **exploring faculty** based on various aspects of their activities, such as *research topics/areas/domains* they work on, *methodologies* they work with, and their current engagements in the *dissertation* committees. Upon graph exploration, a "mid-stage" student who is working on forming a dissertation committee identified potential collaborators they hadn't considered before:

> *Yeah, [the query] narrowed it down a little bit ... But definitely, I would look at [two professors shown in the recall] as being people who are kind of outside of what I was looking at before.*

*Supporting Interdisciplinary Research.* When invited to comment on the utility of the graph, one participant emphasized the importance of its functionality to identify people from different disciplines to collaborate with on **interdisciplinary research**, as monodisciplinary work has its boundaries. To that, a "mid-stage" student said:

> *I want to make an interdisciplinary group on a topic that we might all understand- that's important to me.*

Figure 9 shows the screenshot of a query that represents faculty from different departments (indicated by colors) publishing in the same *domain of application* (i.e., HCI).

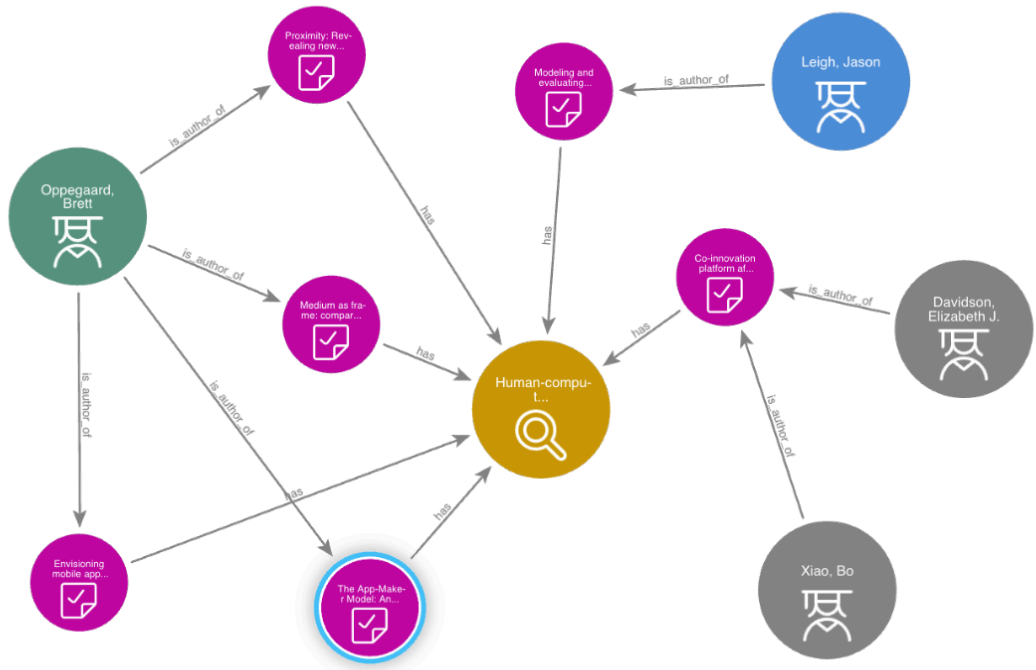

**Figure 9:** Faculty from different departments publishing on Human-Computer Interaction

*Aggregated Data Saves Time*. Three participants reflected on the tool's utility to save time when looking up/searching for the information they need, as the graph aggregates the data from multiple sources to allow for this functionality (Figure 2). Two participants found the controlled vocabularies elicited from this corpus useful to explore categories of their interest. Answering one of the protocol follow-up questions (*Did you find the data you wouldn't be able to find otherwise?),* participants reflected on the hardships of finding not only specific information but rather all the information that was discussed during the interview. An ABD student commented:

> *So, I think a lot of it would be difficult to find, at least all together. Like, you could do this yourself, and it would take forever.*

*Planning for Future Steps Based on Knowledge Graph Insights*

*Decision-Making Support.* Six students expressed the value of the graph, especially when it supports their decision-making process. Codes **helpful** and **decision-making** overlapped five times. When commenting on the criteria for the ratings of queries demonstrated, a "mid-stage" student said:

> *The closer it comes to helping me make a decision, the higher the value it has for me.*

We demonstrated the queries that were voted as useful by PhD students and personalized them to showcase specific categories that were of participants' interest. This helped them to anticipate the next steps in the program. A "new" student interviewee commented on the procedure:

> *This process gave me a lot more information that I didn't even think I needed to know. It answers questions that I didn't know I had.*

Participants found it particularly useful to inform decisions on choosing faculty to collaborate with -as a chair or a committee member ($c=17$). Three students commented on the usefulness of the queries that showed the faculty's current mentoring engagements (example of the query screenshot in Figure 10), as the recall may indicate if they are already "stretched thin" in that segment. One "mid-stage" student participant said:

> *[This query] is very useful for me… I'd want to create a committee with a lot of time for me, right? I think [the professor] working on four dissertations- that's borderline.*

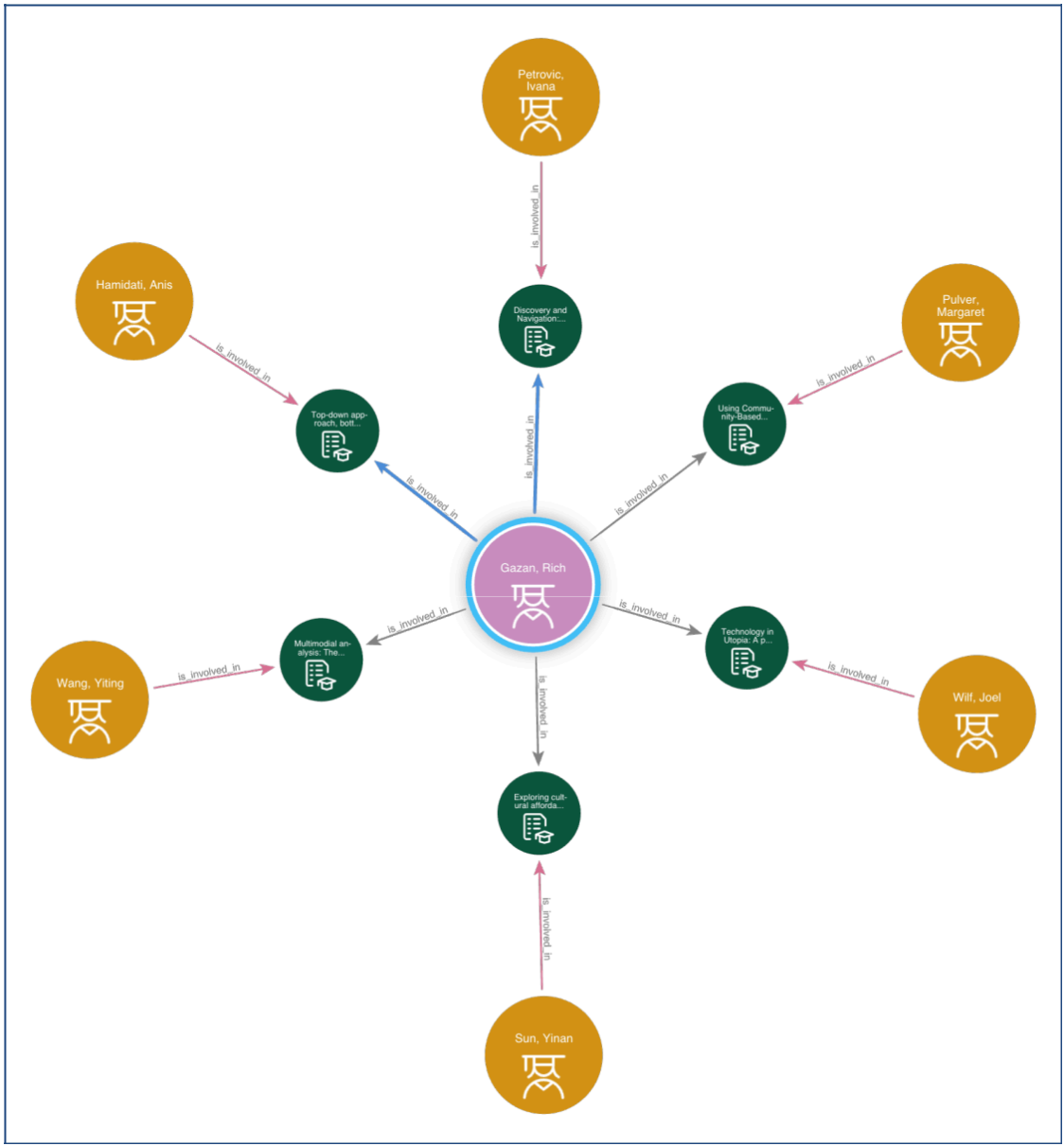

**Figure 10:** Current students of a faculty member (blue edges indicate their chairing role)

*Personal Timeline, Projections, and Reflections.* Throughout the interviews, all participants reflected on their **personal timeline** in the program- not only the time to the degree but also the time to fulfill the set requirements ($c$=42). A "new" student was reminded of their next step:

> *That's just like eye-opening. I didn't even…I honestly kind of forgot about the publishable paper thing. But I guess that's along that line right now.*

Four participants talked about being **timeline-driven**. In such cases, this tool was considered to be very useful regarding their steps planning. A "new"student commented on that:

> *I'm very timeline-driven…so it's really helpful; it gives me a framework because I tend to over-plan, and this puts it in perspective for me.*

The two international students emphasized the usefulness of looking at other students' timelines, as for the international student population, it is helpful to see the "true picture" of the potential length of the program, which helps them with planning around visas, scholarships, and financing. Three participants found the use of the previous students' timelines to compare with them. One of the "mid-stage" student interviewees said:

> *I like to see where I stand amongst my peers. I think it's more important to me because I'm older. And, you know, I'm competing with a bunch of young kids, I guess.*

*Information missing in the Knowledge Graph*
At the end of the evaluation protocol, participants were asked to comment on the information they would like to know that is not represented or can't be represented by the graph. Two participants stated that the graph was comprehensive and captured all the information a CIS Ph.D. student may need. Two participants stated it would be impossible to read negative feedback about professors, as people would refrain from writing about this. An ABD participant said that the graph is a good starting point, yet one needs to do further independent investigations on issues that are not easy to conclude from query recalls (e.g., why is a professor mentoring so few students).

The knowledge graph was noted to be good at showing facts and directing students to people who might provide more insight into events behind the numbers and statistics; to that, a "mid-stage" student commented:

> *I need to talk to each of these people probably, to get that [personalized information].*

Similarly, students were interested in knowing more about one's reasoning behind picking a dissertation supervisor or a committee member. However, when asked if they would leave such a comment themselves, another student laughingly said:

> *Oh, yeah, I mean, I am being sort of hypocritical here. I would love to get that quick and easy information, but you cannot count on me to be a contributor to that.*

We noticed significant differences in the level of comfort when it comes to providing written reflections on faculty and courses between domestic students- who are generally open to it, and international students- who are very much wary of this approach and would avoid the feature altogether. One student talked about the possible misuse of this feature, as it would allow for misinformation-spreading mechanisms in the case of a very competitive program.

Three interviewees were interested to learn more about the personal experiences of students in the program, especially those who quit or were expelled; as well as the professional pathways of alumni (as they follow their degree choices).

**Discussion**

Previous research noted the gaps in the knowledge graph design and evaluation studies, outlining the fact that they are disconnected from the communities they are supposed to serve (Li et al., 2023; McGee et al., 2019). Thus, our study addresses these issues by conducting a qualitative study grounded in participatory approaches for designing the graph, its exploration scenarios, and evaluation of its utility. These approaches were successful in creating a useful database that supports Ph.D. students in information discovery and decision-making, as well as providing potential for information sharing and social connection-building in an interdisciplinary research community. The findings of this study helped us understand the thought process behind interactions with such a tool to improve future knowledge graph iterations and get an idea for the novel multiplex graph interface design.

The implication of this research is especially significant as the methodological contribution to the field, as the knowledge graph research is predominantly addressed via quantitative approach; the human-centered design steps described in this and our previous papers (omited1, omitted2) introduces the iterative development of the graph model and database population that can be replicated to other domains.

*Overall Reactions and Impressions to Knowledge Graph*

The overall impressions of CIS Ph.D. students on the knowledge graph were positive. This finding was evident through the query ratings, comments coding results, and sentiment analysis. Even though previous research has shown that users lose trust in knowledge graphs upon interacting with the complex graph visualizations (Li et al., 2023), the sentiment analysis shows that the participants trust the CIS knowledge graph, which might be due to the choice of queries. The overall acceptance rate and positive feedback for such a tool demonstrate the advantage of the effort made to understand end-user's needs, which are often overlooked when creating knowledge graphs (Li et al., 2023).

The ability to read about the experiences and get ideas on the pathways of other students was appreciated by all participants, although most informative for new students. This is due to the fact that such information is hard to find elsewhere. Ph.D. students found some insights comforting and seemed more confident they would achieve milestones they were initially unclear about. A new student commented on the tool:

> *This is a good navigator; a navigation tool that helps you to know… how to go about it; They need this. This is very handy. Very relevant… I've been struggling with all these things here, and you picked it out.*

It is a promising result, indicating that the designing of a tool to support the activities of Ph.D. students can potentially alleviate some of the stress and strain they are experiencing. Possibly, such an approach might improve student retention by increasing satisfaction with their academic programs, including the fulfillment of their Ph.D. program expectations (Bair and Haworth, 2004).

*Knowledge Graph as a Helpful Tool*
Participants found the tool helpful, especially seeing the connections/links between actors in the domain that one may not have been aware of.

The graph was perceived as helpful as much as it supported the decision-making, in particular when considering which faculty to invite to be their supervisors or to serve as dissertation committee members. This feature is particularly important as we know that many of the challenges and success factors of Ph.D. students are directly related to the student-mentor relationship (Bair and Hagworth, 1999; Grady et al., 2014; Young et al., 2019). The findings emphasize the value of alternative metrics of faculty, such as their mentoring engagements, which are currently not sufficiently recognized when compared to mainstream impact metrics based on publishing (e.g., citation count, h-index, etc.).

Also, the graph was helpful in personal timeline planning and motivation; students could compare their paths to those of their peers.

We will use these findings to inform the features of the knowledge graph interface, emphasizing those that may support intrinsic motivations for contributing and updating crowdsourced data (Pinto and dos Santos, 2018). For this population, those are (i) progress checks and comparisons with other students, (ii) projection of future steps and planning, and (iii) personal graph development through time.

*Missing From the Graph*

All participants agreed that more context is needed to understand better what was happening with people behind the numbers they have seen in the metrics shown. The qualitative component of actual human experience is beyond the ability of any graph to represent.

Sometimes, information may or may not be explicitly encoded in a graph, but students would find value in simply knowing who to ask. The graph makes it easier to identify people with multiple potential shared interests and ask them about their degree choices- for example, why someone chose one supervisor over another. This way, students would be supported in initiating conversations, potentially fostering relational confidence and trusted information exchange (Keppler and Leonardi, 2023), even peer mentorship that is perceived as beneficial for Ph.D. students' satisfaction and retention (Preston et al., 2014).

Although the "previous students timeliness" feature was perceived as beneficial for the study participants to plan for success and get motivated to proceed, the downfalls of having this data available are potential enhancement of the negative feelings caused by imposter syndrome (Chakraverty, 2020) or social comparison (Kim et al., 2021).

**Limitations and Future Work**

The limitations of our approach are mainly about generalizability. Even though the CIS program is small and not representative of the broader Ph.D. education realm, it may be considered a critical example of an interdisciplinary program in which faculty from different departments are affiliated In such programs, PhD students have to navigate various departmental cultures (Gardner, 2011). However, to provide further use of this approach, it can be applied to other interdisciplinary programs, e.g., those offered by iSchools [5]. The iSchool related research has highlighted the potential benefits of a community portal as a forum for peer information exchange among students from various disciplines (Naughton, 2010) and the need to support the identity formation among students

in this interdisciplinary setting, and both can be addressed via this knowledge graph approach (Choi, 2015)

As our approach bears limitations of manual tagging and data input, more research is needed to identify the appropriate mechanism to ingest and update relevant publicly available data, by modifying existing approaches (e.g., Banerjee et al., 2022; Müller, 2023).

Questions of data privacy and transparency must be addressed. For example, students' comments on courses (and faculty mentorship styles) were found to be extremely useful yet potentially harmful features. Also, other stakeholders could be potential users of this knowledge graph, e.g., faculty, administrators, graduate division officers, and alumni. Considering the flexibility of knowledge graph as a tool (Li et al., 2023), such changes could easily be implemented through another round of requirements gathering with the mentioned stakeholders.

**Conclusion**

This paper focused on the evaluation of a knowledge graph designed to support Ph.D. students in their progress toward the degree and their integration into the research and educational community. We have discussed some of the observed advantages and pitfalls of having knowledge and social networks represented in a knowledge graph. We demonstrated that involving end-users in the design process can help students anticipate and plan for the next steps in their degree and potentially connect with one another. The findings have shown that the end users found the tool helpful, especially as it supports them in making data-driven decisions, be more confident of their next steps, and it saves them time when looking for information. Still, the students noted the lack of the qualitative aspect behind the data aggregated in the knowledge graph, and would like to know more of the experiences of the previous students.

This tool can be used for making more personalized recommendations than is possible through generic policies and degree requirements, as well as improving the visibility of the work done by researchers in the local setting. Such an approach overcomes the socio-technical issues that may emerge due to a lack of understanding of end users' needs, but also a methodological deficiency often overlooking the qualitative approach to research by technology developers when designing knowledge graph exploration software.

**Endnotes**

[1] Graph evaluation protocol-

https://drive.google.com/file/d/1FuzWFTBVFYUVSx-uQ0ZVirydo66cw80Z/view

[2] List of controlled vocabularies-

https://docs.google.com/spreadsheets/d/1qgIuqEpebvcDswAC3wzdDPAfCvwx1dIvwLRCqjHKmkE/edit?usp=sharing

[3] The funds were received through the National Science Foundation, under Grant No. 1933803, distributed by the Science of Science Summer School (S4) program, where the co-authors met and started collaborating.

[4] https://otter.ai/

[5] https://www.ischools.org/

**Appendix 1- Voting results on 20 queries on CIS knowledge graph**

The table shows the voting results on the queries (on the five-point Likert scale where five is "most useful"), averaged for *new*, *mid* (second year), and *ABD* students who participated in the evaluation study. The queries are grouped into categories (e.g. *Student pathways*). All of these Cypher queries are available in the published dataset, together with the layout for visualizing them.

| Queries | Student progress in PhD program | | |
|---|---|---|---|
| **Student pathways** | Avg_New | Avg_Mid | Avg_ABD |
| **Q1** How long did it take students to publish a paper? | 4.75 | 5 | 4 |
| **Q2** Students' timeline- to propose and defend | 4.75 | 4.5 | 4.5 |
| **Q3** Venues where students published papers | 4.25 | 4.5 | 5 |
| **Faculty and their dissertation involvements** | Avg_New | Avg_Mid | Avg_ABD |
| **Q4** Number of dissertations all faculty chaired | 3.5 | 4 | 4 |
| **Q5** Number of dissertations faculty were involved in | 4.25 | 4 | 3 |
| **Q6** Number of dissertations Professor X chaired on Area Y | 4 | 3.5 | 5 |
| **Q7** Who are university representatives on dissertations | 3.25 | 3.5 | 4 |
| **Q8** Dissertations faculty chaired on Research Area/ Research Topic/Domain of Application | 3.75 | 4 | 4 |
| **Q9** Faculty dissertation collaborators of Prof X | 3.75 | 4 | 4 |
| **Q10** Current students of Professor X | 4.75 | 5 | 4.5 |
| **Q11** Dissertation committee involvement of Prof X | 4.75 | 5 | 5 |
| **Q12** Dissertation collaborators of Prof X on topic Y | 4.25 | 4 | 3.5 |
| **Courses/exams info** | **Avg_New** | **Avg_Mid** | **Avg_ABD** |
| **Q13** Comments of people who took directed readings with Prof X | 4.75 | 4 | 4 |
| **Q14** Students who took Course/Exam X and their comments | 4.75 | 4 | 4 |
| **Q15** Courses teaching Methodology X, and students who took them | 4.75 | 5 | 4 |
| **Faculty activities (other than *Dissertation*)** | **Avg_New** | **Avg_Mid** | **Avg_ABD** |
| **Q16** Co-authorship network within CIS | 4 | 3.5 | 4.5 |
| **Q17** People who share research topics with Professor X | 4.25 | 3.5 | 4 |
| **Q18** People and publications on Research Area/Topic X | 4.5 | 4.5 | 5 |
| **Q19** Researchers from the different Departments who look at the Domain of Application X | 4 | 4.5 | 4.5 |
| **Q20** Expand the previous query- show Methodologies/Tools they are using | 3.25 | 4.5 | 5 |

**Table 1:** CIS KG Queries and their Ratings

**Appendix 2: Step-by-Step Procedure to Perform Sentiment Analysis in R**

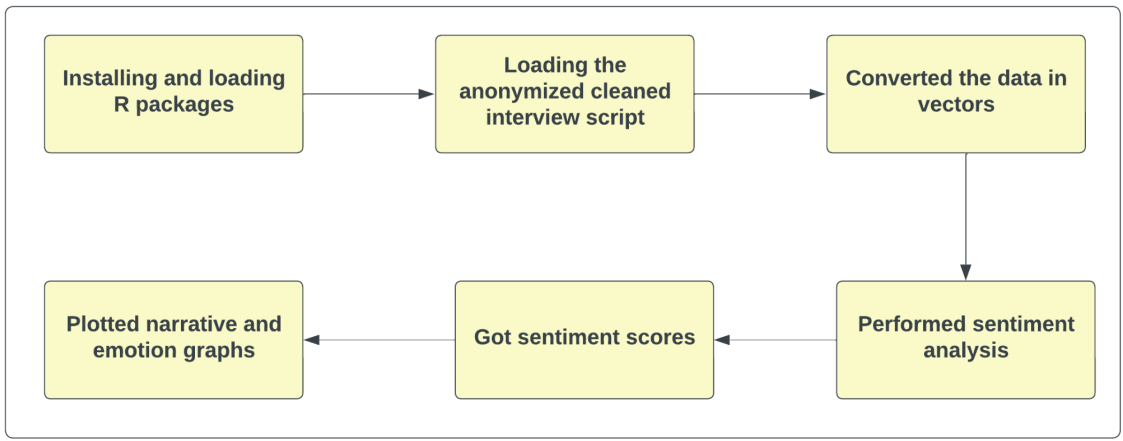